\documentclass[a4paper,11pt]{article}
\usepackage{pos}
\usepackage{diagbox}
\usepackage{makecell, multirow}
\usepackage{amsmath}
\usepackage{braket}
\usepackage{subcaption}
\usepackage[percent]{overpic}
\allowdisplaybreaks
\definecolor{maroon}{RGB}{102,0,51}

\newcommand{\ala}{{\it \'a la}}

\title{Doubly Bottom and Bottom-Strange Tetraquarks in the Isoscalar Channel}

\author*[1,3]{Bhabani Sankar Tripathy}
\author[2]{Nilmani Mathur}
\author[1,3]{M.~Padmanath}

\affiliation[1]{The Institute of Mathematical Sciences,  CIT Campus, Chennai, 600113, India}
\affiliation[2]{Department of Theoretical Physics, Tata Institute of Fundamental Research, Homi Bhabha Road, Colaba, Mumbai 400005, India}
\affiliation[3]{Homi Bhabha National Institute, Training School Complex, Anushaktinagar, Mumbai 400094, India}

\emailAdd{bhabanist@imsc.res.in}

\FullConference{The 42nd International Symposium on Lattice Field Theory (LATTICE2025)\\
	2-8 November 2025\\
	Tata Institute of Fundamental Research, Mumbai, India\\}

\abstract{We present our recent investigation on doubly bottom and bottom-strange tetraquarks in the isoscalar channel in search of a possible tetraquark bound state. The calculations are performed on four ensembles with dynamical quark fields up to the charm quark generated by the MILC Collaboration with various lattice spacings. Two volumes have been used to account for finite volume effects. Overlap action has been employed to calculate light and strange quark propagators and NRQCD formulation is utilized for heavy bottom quarks. Finite volume energy has been calculated using the variational method followed by rigorous scattering amplitude analysis à la Lüscher. We find strong evidence for a deeply bound state in the doubly bottom tetraquark channel, but no conclusive evidence for the existence of a bottom–strange tetraquark.}

\begin{document}
\maketitle

\section{\label{sec:intro}Introduction and Motivation}
Murray Gell-Mann, in his seminal paper~\cite{Gell-Mann:1964ewy}, described the known conventional hadrons—namely three-quark baryons and quark–antiquark mesons—and also indicated the possibility of more complicated structures such as tetraquarks and pentaquarks. For many years, clear experimental evidence for such states was lacking. Over the past two decades, numerous unconventional hadrons often referred to as the XYZ states have been observed, particularly in the heavy-quark sector~\cite{Chen:2022asf,Brambilla:2019esw,Brambilla:2022ura}. Prominent examples include the $X(3872)$, $Z_c$, $Z_b$, and $P_c$ states. A major development was the discovery of the $T_{cc}^+$ tetraquark by the LHCb collaboration~\cite{LHCb:2021vvq,LHCb:2021auc}, which carries four distinct valence flavors and lies near the $DD^*$ threshold. This state belongs to the class of doubly heavy tetraquarks, systems that are expected to become stable in the heavy-quark mass limit based on heavy-quark symmetry arguments~\cite{Carlson:1987hh,Manohar:1992nd,Eichten:2017ffp}.

An important open question concerns whether the bottom quark is sufficiently heavy to guarantee binding in the $bb\bar{u}\bar{d}$ channel below the $BB^*$ threshold. A wide range of phenomenological approaches including quark models, QCD sum rules, effective field theory analyses, and chromomagnetic interaction models have addressed this issue and generally support the existence of a strongly bound doubly-bottom tetraquark. Nevertheless, experimental observation of such states remains challenging due to the large center-of-mass energies required to produce two bottom quarks. Possible detection strategies for bottomness-$2$ tetraquarks have been discussed in Refs.~\cite{Moinester:1995fk,Ali:2018ifm,Ali:2018xfq}.

Lattice QCD provides a first-principles tool for studying exotic hadrons. Early calculations explored $bb\bar{u}\bar{d}$ tetraquarks with considering bottom quarks in the static limit \cite{Bicudo:2012qt,Bicudo:2015vta}, while recent simulations incorporate dynamical light/strange quarks and treat bottom quarks via NRQCD formalism \cite{Francis:2016hui,Junnarkar:2018twb,Leskovec:2019ioa,Mohanta:2020eed,Hudspith:2023loy,Aoki:2023nzp,Alexandrou:2024iwi,Colquhoun:2024jzh}. Our earlier work also used NRQCD methods to treat the evolution of b-quark \cite{Junnarkar:2018twb}. These studies consistently find a strong-interaction-stable axialvector $bb\bar{u}\bar{d}$ tetraquark. Investigations have extended to $bc\bar{u}\bar{d}$ \cite{Padmanath:2023rdu,Radhakrishnan:2024ihu} and $bb\bar{u}\bar{s}$ \cite{Leskovec:2019ioa,Hudspith:2023loy}. 

In this work, we investigate tetraquark systems containing bottom quarks, specifically focusing on the $bb\bar{u}\bar{d}$~($T_{bb}$) tetraquark with quantum numbers $I(J^P)=0(1^+)$ and the $bs\bar{u}\bar{d}$~($T_{bs}$) tetraquark with quantum numbers $I(J^P)=0(1^+)$ and $0(0^+)$. The study involves the use of four different ensembles with two different spatial volumes, facilitating a finite-volume scattering analysis. Such a finite-volume analysis was not performed in our previous work \cite{Junnarkar:2018twb}. The study of isoscalar $bs\bar{u}\bar{d}$ channels is a extension of our previous calculations of a similar tetraquark system with flavor content $bc\bar u\bar d$ \cite{Padmanath:2023rdu, Radhakrishnan:2024ihu}. 

In this proceedings contribution, we present our investigation of doubly-bottom and bottom-strange tetraquarks in the isoscalar sector. The remainder of this article is organized as follows. In Sec.~\ref{sec:setup}, we describe the lattice setup employed in this study. Section~\ref{sec:currents} provides a brief discussion of the interpolating operators used in our analysis. Our results are presented in Sec.~\ref{sec:results}, and finally, we summarize our conclusions in Sec.~\ref{sec:conc}.
\section{\label{sec:setup}Numerical Setup}
We worked with four ensembles of lattice QCD gauge configurations generated by the MILC collaboration \cite{MILC:2012znn}, featuring $N_f = 2+1+1$ dynamical fermion flavors simulated using the Highly Improved Staggered Quark (HISQ) action. These ensembles were specifically chosen to investigate the volume and lattice spacing dependence of our results. The finest lattice spacing employed is $a \sim 0.0582$ fm. The dynamical  strange and charm quarks were tuned to their physical values, while the light quarks were simulated with heavier pion masses in isospin symmetric limit. Detailed ensemble parameters are provided in Table~1 of Ref.~\cite{Tripathy:2025vao}.

For the valence sector, we employed overlap fermions. Given the computational expense of overlap fermions and the prohibitively large statistical noise encountered when tuning light quark masses below 0.5 GeV, we instead investigated the light quark mass dependence using unphysical pseudoscalar meson masses of 0.5, 0.6, 0.7, and 3.0 GeV. The propagators generated with $M_{ps} \sim 0.7$ GeV and $\sim 3.0$ GeV correspond to the physical strange quark (688.5 MeV) and the charm quark (using the kinetic mass determined via the Fermilab prescription), respectively. We employ non-relativistic QCD to describe bottom quark evolution Ref \cite{Lepage:1992tx}.

\section{\label{sec:currents}Interpolating Operators}
The low-lying hadron spectrum is extracted from the time dependence of two-point current-current correlation matrices in Euclidean spacetime, whose elements are given by:
\begin{equation}
\mathcal{C}_{ij}(t) = \sum_{\mathbf{x}}\langle\Phi_i(\mathbf{x},t)\tilde \Phi_j^{\dagger}(0)\rangle = \sum_n \frac{Z_i^n\mathcal{Z}_j^{n\dagger}}{2E^n} e^{-E^nt}.
\label{twoptc}
\end{equation}

Here, ${\Phi_i(\mathbf{x},t)}$ represents a carefully constructed basis of interpolating currents with the quantum numbers corresponding to the state of interest. The mass or energy of the state can be extracted from the large-time behavior of these correlation functions. The operator-state overlap strength $Z_i^n = \braket{0|\Phi_i|n}$ determines the coupling of the interpolator $\Phi_i$ to the state $n$.

We employed wall source smearing for the quarks and point sink for optimal overlap with ground states. However this asymmetric setup leads to correlation functions that approach the large time plateau in effective mass from below, potentially resulting in the large time plateau misidentification for ground state. We overcome this issue through a comparative study employing box sink smearing \cite{Hudspith:2020tdf}, where increasing the box radius progressively restores symmetry, ultimately yielding a wall source-wall sink correlation function. For all three systems, we employed local meson-meson and local diquark-antidiquark interpolating operators.

The lowest relevant two-body scattering channel for the isoscalar axialvector $T_{bb}$ corresponds to $BB^*$, while the lowest inelastic scattering channel is $B^*B^*$. For the isoscalar axialvector $bb\bar{u}\bar{d}$ tetraquark, we utilize the same set of interpolators as in Ref.~\cite{Junnarkar:2018twb}. We consider two operators: one associated with the $BB^*$ channel and another of the diquark-antidiquark form, given below.
	\begin{align}
		\Phi_{\mathcal{M}_{BB^*}}(x) &= \left[ \bar{u}(x)\gamma_i b(x)\right] \left[ \bar{d}(x)\gamma_5 b(x)\right] 
		 - \left[ \bar{u}(x)\gamma_5 b(x)\right] \left[ \bar{d}(x)\gamma_i b(x)\right], &\nonumber \\
		\Phi_\mathcal{D}(x) &= \big[ \left(\bar{u}(x)^T C\gamma_5\bar{d}(x) - \bar{d}(x)^T C\gamma_5\bar{u}(x)\right) 
		\times (b^T(x)C\gamma_i b(x))\big].
		\label{eq:bb1}
	\end{align}
For the isoscalar axialvector $T_{bs}$, our focus is on $S$-wave $K\bar{B}^*$ scattering in the rest frame, which yields infinite-volume quantum numbers $I(J^P)=0(1^+)$ and reduces to the $T_1^+$ finite-volume irrep. We employ an operator basis similar to that used in Ref.~\cite{Padmanath:2023rdu} for the study of isoscalar axialvector $bc\bar u\bar d$ tetraquarks, with the charm quark replaced by a strange quark. The operator basis is given below. 
\begin{align}
		\Phi_{\mathcal{M}_{KB^*}}(x) &= \left[ \bar{u}(x)\gamma_i b(x)\right] \left[ \bar{d}(x)\gamma_5 s(x)\right] 
		- \left[ \bar{u}(x)\gamma_5 s(x)\right] \left[ \bar{d}(x)\gamma_i b(x)\right], \nonumber \\
		\Phi_{\mathcal{M}_{BK^*}}(x) &= \left[ \bar{u}(x)\gamma_5 b(x)\right] \left[ \bar{d}(x)\gamma_i s(x)\right] 
		- \left[ \bar{u}(x)\gamma_i s(x)\right] \left[ \bar{d}(x)\gamma_5 b(x)\right], \nonumber \\
		%
		\Phi_\mathcal{D}(x) &= \big[ 
		\left(\bar{u}(x)^T C\gamma_5 \bar{d}(x) - \bar{d}(x)^T C\gamma_5 \bar{u}(x)\right) 
		 \times \left(b^T(x) C\gamma_i s(x)\right)
		\big] .
		\label{eq:bs1}
	\end{align}

For the isoscalar scalar $T_{bs}$, the lowest relevant two-particle scattering threshold is associated with the $K\bar B$ channel in $S$-wave. We employ the two-operator basis used in Ref.~\cite{Radhakrishnan:2024ihu} for the study of isoscalar scalar $bc\bar u\bar d$ tetraquarks, with the charm quark replaced by a strange quark. The operators are listed below.
\begin{align}
		\Phi_{\mathcal{M}_{BK}}(x) &= \left[ \bar{u}(x)\gamma_5 b(x)\right] \left[ \bar{d}(x)\gamma_5 s(x)\right] \
		- \left[ \bar{u}(x)\gamma_5 b(x)\right] \left[ \bar{d}(x)\gamma_5 s(x)\right], \nonumber \\
		%
		\Phi_\mathcal{D}(x) &= \big[ \left(\bar{u}(x)^T C\gamma_5\bar{d}(x) - \bar{d}(x)^T C\gamma_5\bar{u}(x)\right) 
		\times (b^T(x)C\gamma_5 s(x))\big].
		\label{eq:bs0}
\end{align}

In Eqs.~\ref{eq:bb1}, \ref{eq:bs1}, and \ref{eq:bs0}, $C = i\gamma_y\gamma_t$ represents the charge conjugation matrix. For all interpolating operators specified above, the quantities inside the square brackets are color singlets.
\section{\label{sec:results}Results}
We construct correlation matrices and analyze them variationally using the generalized eigenvalue problem (GEVP):
\begin{equation}
		\mathcal{C}(t)v^{(n)}(t) = \lambda^{(n)}(t) \mathcal{C}(t_0)v^{(n)}(t).
		\label{gevp}
\end{equation}
evaluated at each time slice. Following the GEVP, we obtain correlation functions corresponding to the ground and excited states of interest. It is important to note that the wall-source smearing setup was chosen to achieve optimal overlap with ground states, which we confirmed through extended plateaus. This approach provided confidence in fitting the correlators, albeit at the cost of limited access to excited states, as discussed in Ref.~\cite{Padmanath:2023rdu}. Energies were extracted by fitting the leading exponential obtained from the solution of the GEVP.
\begin{figure}[htbp]
\centering
\begin{minipage}{0.48\linewidth}
    \centering
    \includegraphics[width=\linewidth,height=3.8cm]{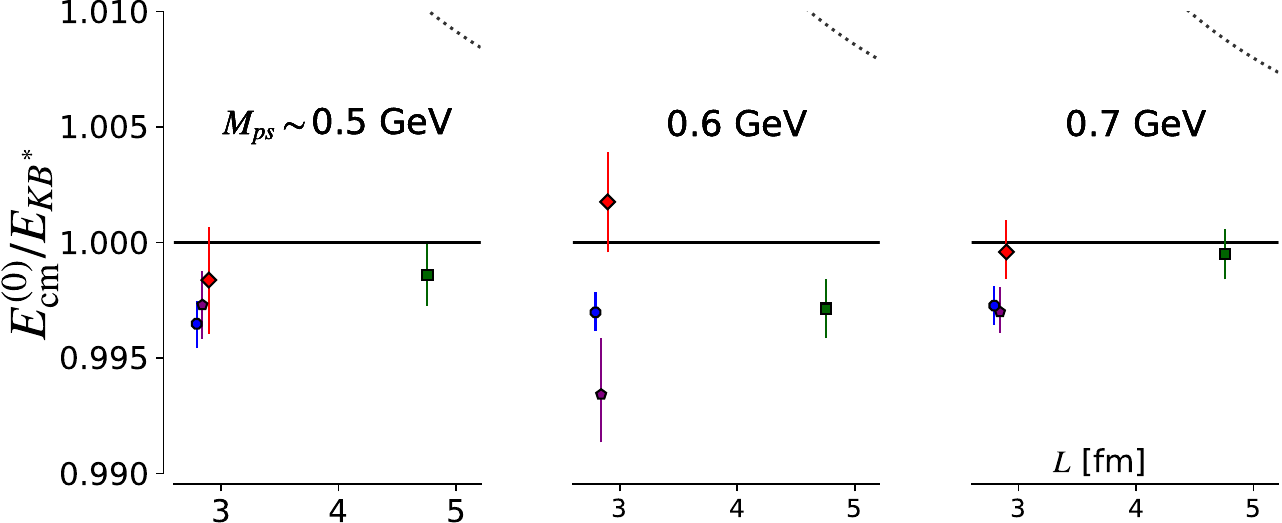}
\end{minipage}
\hfill
\begin{minipage}{0.48\linewidth}
    \centering
    \includegraphics[width=\linewidth,height=3.8cm]{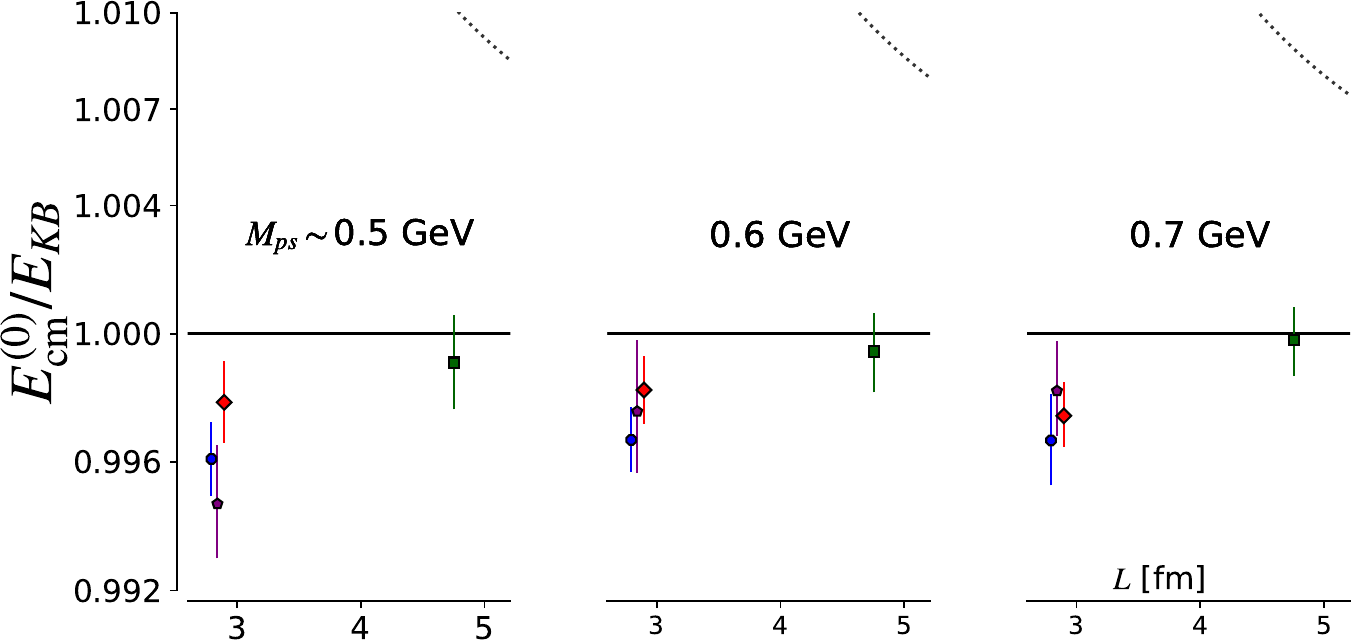}
\end{minipage}

\vspace{0.5em}

\begin{minipage}{0.6\linewidth}
    \centering
    \includegraphics[width=\linewidth]{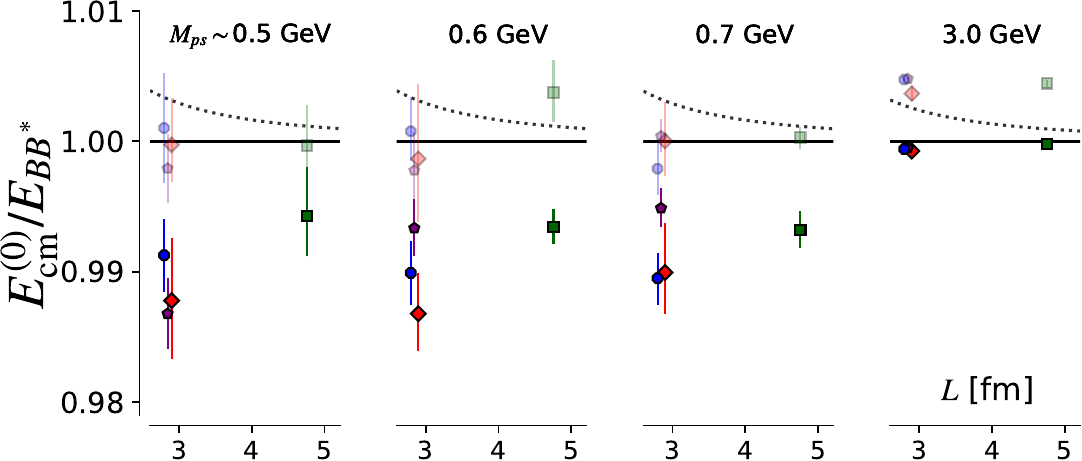}
\end{minipage}

\caption{Ground-state energy spectra of the $T_{bs}$ system are shown in the top two panels. The bottom panel shows the ground and excited state energy spectrum of the $T_{bb}$ tetraquark. The excited state of $T_{bb}$ is indicated by faded markers. All energies are expressed in units of the corresponding two-meson decay thresholds, and the horizontal axis denotes the spatial lattice extent.}
\label{fig:spectrum}
\end{figure}
In Figure~\ref{fig:spectrum}, we present the ground state energies of the axialvector $T_{bb}$, axialvector $T_{bs}$, and scalar $T_{bs}$ systems. The spectra are shown in units of their respective lowest two-meson decay thresholds. The energy values presented in Figure~\ref{fig:spectrum} are free from additive corrections inherent to hadron correlators containing bottom valance quark realized using NRQCD action \cite{Padmanath:2023rdu, Tripathy:2025vao}.

For $T_{bb}$, negative energy shifts in the ground states relative to the elastic $BB^*$ threshold are evident across all light quark mass cases studied, suggesting attractive interactions between the $B$ and $B^*$ mesons. Another observable effect is the gradual decrease in binding energy (in units of the elastic threshold energy) with increasing pseudoscalar meson mass $M_{ps}$. This indicates a weakening of the attractive interactions as the light quark mass increases, implying that the attraction becomes stronger for lighter light quarks or heavier heavy quarks.

For the axialvector $T_{bs}$ channel, we present the finite-volume ground state energy estimates in units of the elastic threshold $E_{KB^*}$. No significant variation in the energy shifts relative to the threshold is observed as a function of $M_{ps}$. The green square corresponding to the large-volume ensemble clearly suggests energy estimates consistent with the threshold across all $M_{ps}$ values, indicating negligible interactions, if any. A similar behavior is observed for the scalar $T_{bs}$ channel, where the finite-volume energy is presented in units of the nearest two-meson strong decay threshold.

Following the extraction of ground state energies, we extract the $S$-wave amplitudes in the channels studied within the elastic approximation, following the finite-volume two-particle spectrum quantization prescription \ala~L\"uscher and its generalizations \cite{Luscher:1990ux, Briceno:2014oea}. For all three cases investigated, under the assumption of elastic scattering, the $S$-wave phase shift $\delta_{l=0}(k)$ can be extracted from the finite-volume spectral energies via the quantization relation:
\begin{equation}
p\cot[\delta_0(p)] = \frac{2}{\sqrt{\pi}L} Z_{00}\left[1;\left(\frac{pL}{2\pi}\right)^2\right],
\end{equation}

where $p$ is the momentum of the scattering particles in their center-of-momentum frame. The momentum $p$ is related to the total energy $E_{cm}$ through:
\begin{equation}
4sp^2 = \left(s-(M_{1}+M_{2})^2\right)\left(s-(M_{1}-M_{2})^2\right),
\end{equation}
with $s=E_{cm}^2$ the Mandelstam variable and $M_i$ the mass of each scattering particle.

We  performed the L\"uscher-based amplitude fit using the finite-volume $BB^*$ scattering data, assuming a zero-range approximation for the amplitude and incorporating a linear dependence on the lattice spacing $a$ to account for cutoff effects. Under this assumption: 
	\begin{equation}
	    p~\cot(\delta_0) = A^{[0]}+a \cdot A^{[1]},
	\end{equation}\label{zerorange}
where $A^{[0]} = -1/a_0$ represents the negative inverse scattering length in the continuum limit.
\begin{figure}[htbp]
\centering
\begin{minipage}{0.48\linewidth}
    \centering
    \includegraphics[width=\linewidth,height=3.2cm]{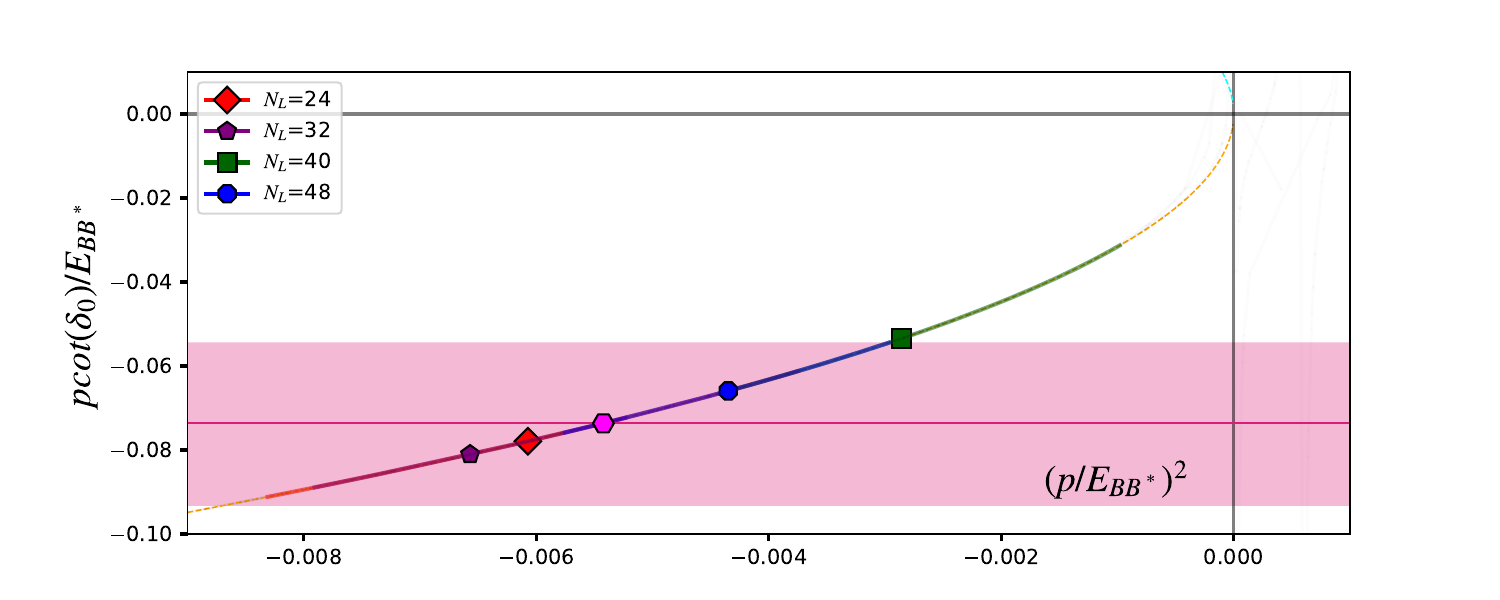}
\end{minipage}
\hfill
\begin{minipage}{0.48\linewidth}
    \centering
    \includegraphics[width=\linewidth,height=3.2cm]{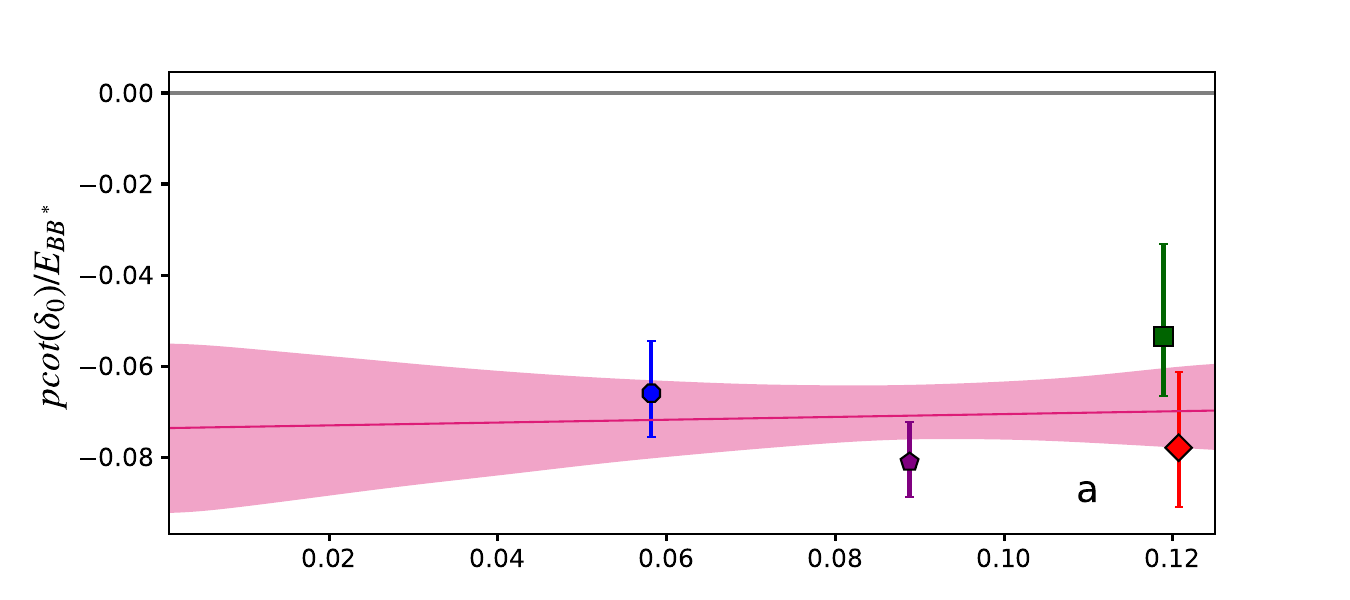}
\end{minipage}
\vspace{0.5em}
\begin{minipage}{0.6\linewidth}
    \centering
    \includegraphics[width=\linewidth]{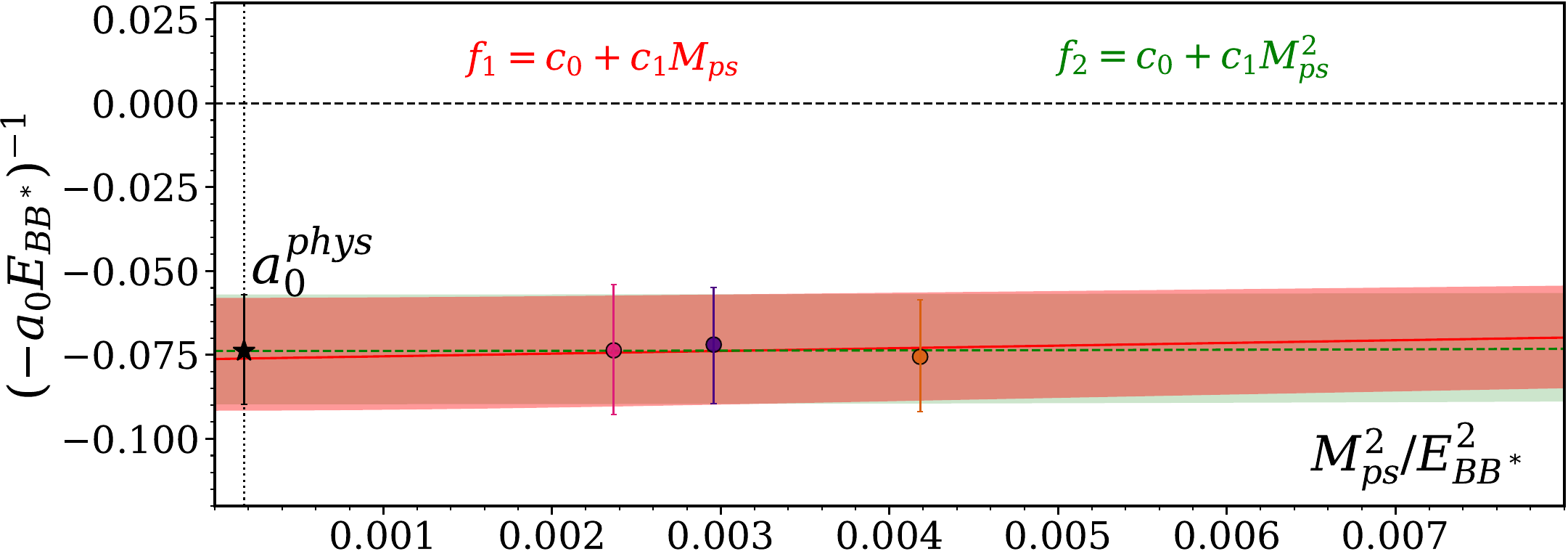}
\end{minipage}
\caption{Upper left: Energy dependence of $p\cot\delta$ normalized to the threshold for $M_{ps}\sim 0.5$ GeV. Upper right: Lattice spacing dependence of $p\cot\delta$ in terms of the threshold for $M_{ps}\sim 0.5$ GeV. Lower: Chiral extrapolation of amplitudes calculated in the continuum limit. The star symbol at the physical pion mass represents the amplitude at the physical point $(a_0^{\text{phys}}E_{BB^*})^{-1}$.}
\label{fig:bbud}
\end{figure}
A particle existing in nature corresponds to a pole in the complex energy plane, as illustrated in the upper left subfigure of Figure~\ref{fig:bbud} for the case $M_{ps}\sim 0.5$ GeV. The magenta pentagon marker indicates the point where the parametrization $p\cot(\delta_0)$ in the continuum limit intersects the constraint curve $ip$. The resulting pole position suggests a real bound state. We performed similar calculations for other $M_{ps}$ values and obtained real bound states in every case. The upper right subfigure shows the continuum extrapolation of $p\cot(\delta_0)$ following Eq.~\ref{zerorange} as a function of the lattice spacing $a$. Similar continuum extrapolations were performed for other $M_{ps}$ cases, yielding negative amplitudes in all instances. Error bars represent $1\sigma$ uncertainties. Using the continuum-extrapolated data for $M_{ps} \sim 0.5,~0.6,~0.7$ GeV, we performed a chiral extrapolation using the fit form $f(M_{ps}^2) = c_0 + c_1 M_{ps}^2$, motivated by chiral effective field theory. The resulting chiral extrapolated amplitude indicates a bound state pole in the physical $BB^*$ amplitude with a binding energy of $\Delta E_{T_{bb}}(1^+) = -116(^{+30}_{-36})$ MeV.

\begin{figure}[htbp]
\centering
\begin{minipage}{0.48\linewidth}
    \centering
    \includegraphics[width=\linewidth]{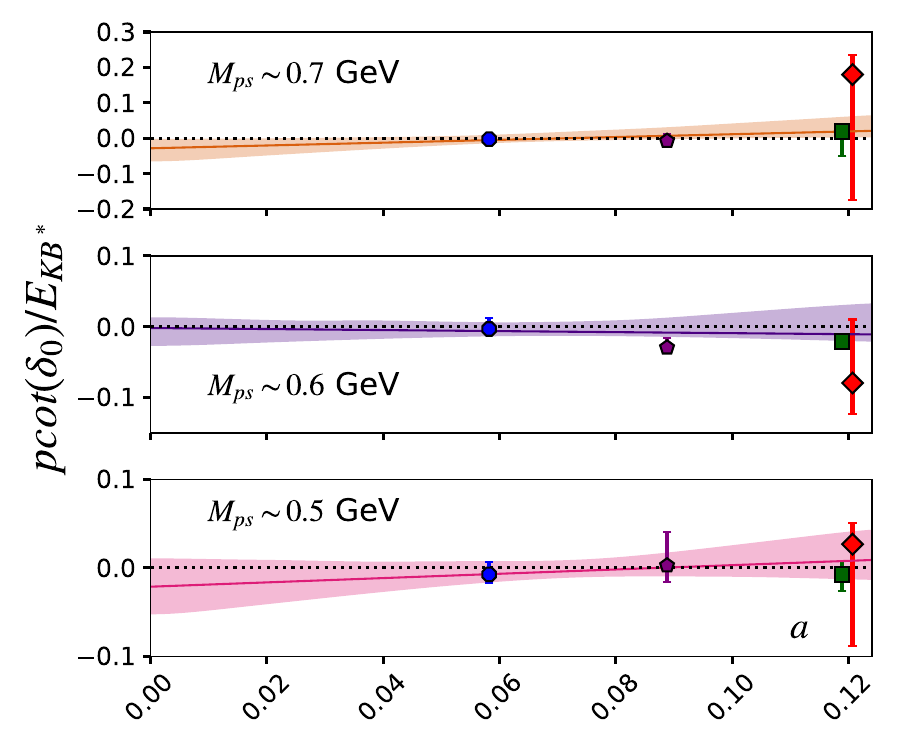}
\end{minipage}
\hfill
\begin{minipage}{0.48\linewidth}
    \centering
    \includegraphics[width=\linewidth]{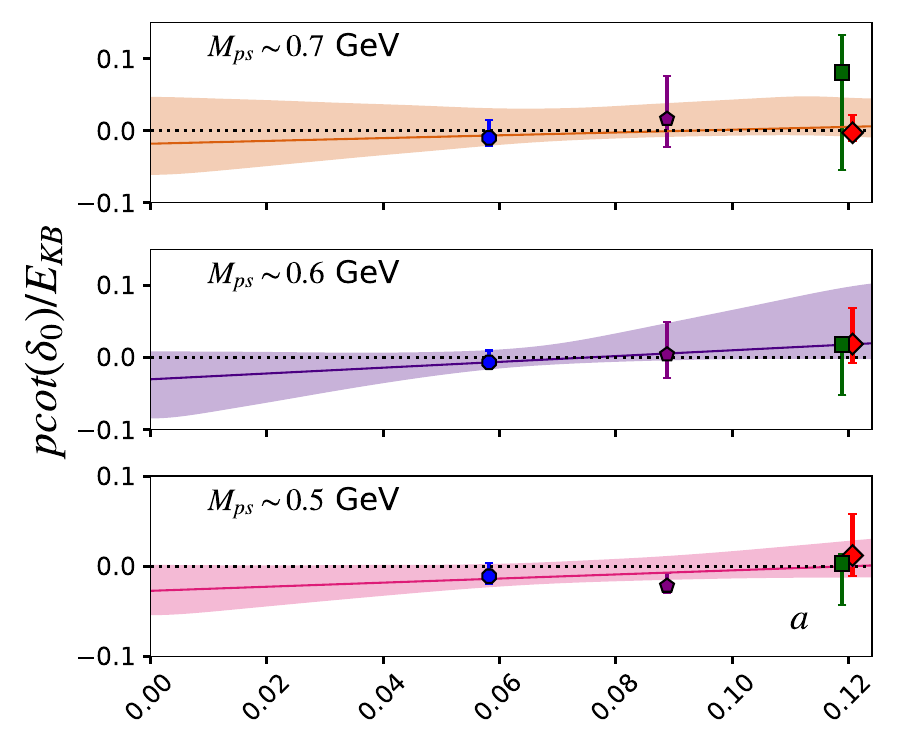}
\end{minipage}
\caption{Continuum-extrapolated results for the axialvector (left) and scalar (right) $bs\bar{u}\bar{d}$ tetraquarks, presented in units of their respective decay thresholds $B^*K$ and $BK$. The colored bands indicate $1\sigma$ uncertainties.}
\label{fig:bsud}
\end{figure}
In Figure~\ref{fig:bsud}, we present the L\"uscher-based amplitude fit results for the finite-volume $B^*K$ and $BK$ scattering data. We again assume the zero-range approximation for the amplitude and employ a linear fit form in $a$ following Eq.~\ref{zerorange}. From both panels, it is evident that the continuum-extrapolated $p\cot(\delta_0)$ values are consistent with zero across all $M_{ps}$ values studied. It can also be observed that given the large uncertainties in $p\cot(\delta_0)$, any estimates of continuum binding energies would be similarly smeared out with substantial uncertainties. These results indicate that the system does not support any bound states.

\section{\label{sec:conc}Summary and Outlook}
We performed a lattice QCD investigation of isoscalar tetraquark channels with bottom quarks on four HISQ lattice ensembles generated by the MILC collaboration with $N_f=2+1+1$. The valence quark dynamics are studied using an overlap formulation of the lattice fermion action for the quark masses up to the charm quark mass, whereas an NRQCD formulation was utilized to describe the bottom quark evolution. Special emphasis were given on using multiple ensemble with varying spatial volume and lattice spacing to control potential discretization effects. 

This study addresses three isoscalar tetraquark channels such as axialvector $bb\bar u\bar d$ and axialvector as well as scalar $bs\bar u\bar d$ tetraquarks. In all of the cases we  investigate the low energy finite-volume eigenenergies across the ensembles, followed by a finite-volume scattering analysis, \ala~L\"uscher, to estimate the binding energy. Along the way, we also employ recently proposed box-sink smearing procedures to identify the ground state saturation in the correlators we utilize.
	
For each of the three channels studied, we employed two types of operator bases: local two-meson interpolators corresponding to the lowest-lying two-meson thresholds, and local diquark–antidiquark interpolators. Ground state energies were extracted via a GEVP analysis. In the $bb\bar u\bar d$ sector, the finite-volume ground state energies exhibit negative shifts relative to the $BB^*$ threshold, indicating a potential $T_{bb}$ tetraquark bound state with binding energy $\Delta E_{T_{bb}}(1^+)=-116(^{+30}_{-36})$ MeV. This estimate, derived from a finite-volume analysis that accounts for lattice spacing dependence, shows no significant sensitivity to the light pseudoscalar meson mass $M_{ps}$. In contrast, the finite-volume ground state energies in the $bs\bar u\bar d$ sector reveal no statistically significant evidence for bound states in either the axialvector or scalar channels.\\
\begin{figure}[htbp]
\centering
\begin{overpic}[width=\linewidth]{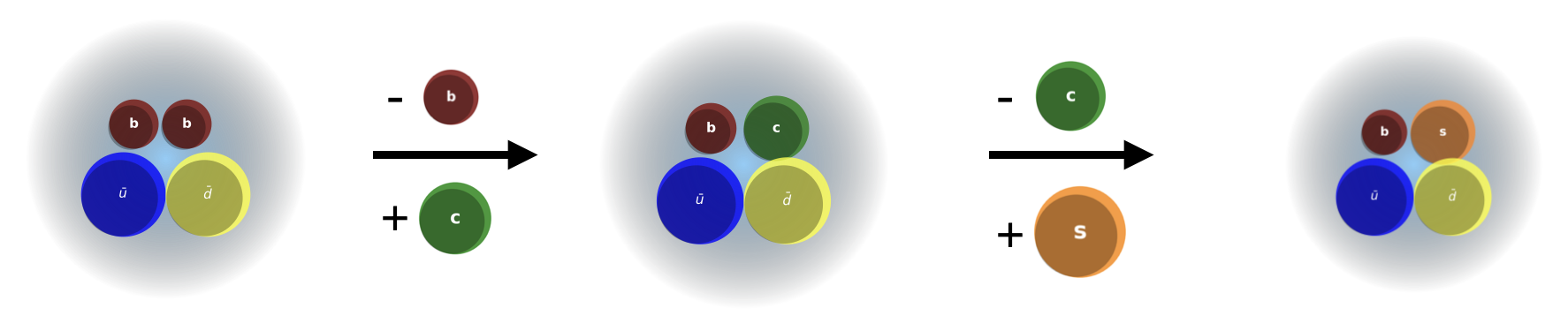}
    \put(3,-1){$\Delta E \sim -116~\mathrm{MeV}$}
    \put(40,-1){$\Delta E \sim -40~\mathrm{MeV}$}
    \put(83,-1){Weakly bound}
\end{overpic}
\caption{Mass dependence of the $bb\bar{u}\bar{d}$ tetraquark binding as one bottom quark is replaced by a charm quark and subsequently by a strange quark. The binding energy decreases in magnitude as one of the heavy quark mass is replaced with light quark, becoming weakly bound in the strange sector.}
\label{fig:quark_mass}
\end{figure}
Finally, we provide a phenomenological interpretation of the binding pattern observed for the $bb\bar{u}\bar{d}$ and $bs\bar{u}\bar{d}$ systems studied in this work, as well as for the $bc\bar{u}\bar{d}$ state reported in Refs.~\cite{Padmanath:2023rdu,Radhakrishnan:2024ihu}. For heavy–heavy light–light tetraquark systems, the spin–spin interaction plays an important role in determining the binding. The strength of this interaction scales inversely with the heavy-quark mass (more precisely the reduced mass of the heavy diquark) which can be explained via heavy quark spin symmetry, heavy antidiquark-diquark symmetry or chromomagnetic interaction models \cite{Eichten:2017ffp, Song:2023izj,Deng:2018kly, Guo:2021yws}. Consequently, for very heavy quarks such as bottom, the spin–spin interaction is strongly suppressed. In the $bb\bar{u}\bar{d}$ system, the spin–spin contribution is therefore negligible, resulting in minimal short-range repulsion and allowing the attractive dynamics to generate a deeply bound state with binding energy of order $\mathcal{O}(100)\,\mathrm{MeV}$. When one of the bottom quarks is replaced by a charm quark, the effective heavy-quark mass scale decreases. As a result, the spin–spin interaction becomes more significant, increasing the repulsive contribution and reducing the overall binding to $\mathcal{O}(40)\,\mathrm{MeV}$. Finally, in the $bs\bar{u}\bar{d}$ system, the bottom–strange heavy-quark pair has a much smaller reduced mass compared to the $bb$ and $bc$ cases. This enhances the spin–spin interaction, increasing the repulsive contribution which leads to a weakly binding in nature. This qualitative behavior is illustrated in Fig.~\ref{fig:quark_mass}.

\vspace{20 pt} 
\acknowledgments  
This work is supported by the Department of Atomic Energy, Government of India, under Project Identification Number RTI 4012. MP gratefully acknowledges support from the Department of Science and Technology, India, SERB Start-up Research Grant No. SRG/2023/001235. We are thankful to the MILC collaboration and, in particular, to S. Gottlieb for providing us with the HISQ lattice ensembles. Computations were carried out on the Cray-XC30 of ILGTI, TIFR, and the computing clusters at DTP, TIFR Mumbai, and IMSc Chennai.


\bibliographystyle{JHEP}
\bibliography{proc}
\end{document}